\documentclass[12pt]{article}
\usepackage{graphicx}
\newcommand{\bb}{\begin{equation}}
\newcommand{\ee}{\end{equation}}
\newcommand{\ba}{\begin{eqnarray}}
\newcommand{\ea}{\end{eqnarray}}

\begin{document}

\title{{\bf Graviton Number Radiated\\
During Binary Inspiral}}

\author{
Don N. Page
\thanks{Internet address:
profdonpage@gmail.com}
\\
Department of Physics\\
4-183 CCIS\\
University of Alberta\\
Edmonton, Alberta T6G 2E1\\
Canada
}

\date{2024 September 25}

\maketitle
\large
\begin{abstract}
\baselineskip 20 pt

The total number of gravitons emitted during nonrelativistic inspiral of two black holes or other effectively point masses is calculated approximately and found to be remarkably close to (well within 1\% of, perhaps within about 0.2\% of) the initial orbital angular momentum divided by the spin angular momentum of each graviton ($2\hbar$), although at unit initial eccentricity, $2\hbar$ times the initial graviton number emission per angular momentum emission is $248/(45\sqrt{3}\pi) \approx 1.012\,811\,600\,479$, almost 1.3\% greater than unity. 

\end{abstract}

\normalsize

\baselineskip 22 pt

\newpage

\section{Introduction}

In recent papers I have pointed out that discrete orbit effects invalidate the previously widely used adiabatic approximation for the evolution of the nonrelativistic energy, semimajor axis, and period of inspiraling binary black holes \cite{Page:2024rsv}, and I have calculated `exact' (in the nonrelativistic approximation) and approximate formulas for the inspiral time as functions of the black hole masses, initial relative velocity, and impact parameter, as well as approximate inverse functions for the impact parameter in terms of the inspiral time and other parameters \cite{Page:2024zpr}.

Here I wish to give an `exact' formula for the number of gravitons emitted per orbital angular momentum lost by highly eccentric nonrelativistic orbits, and approximate formulas for that ratio during the entire inspiral from a given initial eccentricity, as well as for the total number of gravitons emitted during the entire inspiral.

The result may be briefly summarized by noting that the number of gravitons emitted is slightly more than the loss of orbital angular momentum divided by $2\hbar$, the spin angular momentum of a graviton, but the excess is never much more than 0.2\% for the entire inspiral emission of gravitons and angular momentum.

\section{Infinite Series for Graviton Number Emitted}

Consider two black holes (or other effectively point masses, each much smaller than their separation distance) with masses $M_1$ and $M_2$, total mass $M = M_1 + M_2$, reduced mass $\mu = M_1 M_2/M$, and dimensionless mass ratio $\eta = M_1M_2/(M_1+M_2)^2 = \mu/M$, and assume they are slowly inspiraling along a sequence of nearly Keplerian orbits, each with semimajor axis $a$, eccentricity $e$, period $\tau = 2\pi\sqrt{a^3/GM}$, nonrelativistic energy $E = -GM^2\eta/(2a)$, and angular momentum $L = \sqrt{GM^3\eta^2 a(1-e^2)}$.

Calculations by Philip Carl Peters and by Peters and Jon Mathews
\cite{Peters:1963ux,Peters:1964qza,Peters:1964zz} show that the average gravitational power per orbit radiated into the $n$th harmonic (angular frequency $\omega = n\sqrt{GM/a^3}$) is
\bb
P(n)=\frac{32}{5}\frac{G^4}{c^5}\frac{M^5\eta^2}{a^5}g(n,e),
\label{Pn}
\ee
\ba
g(n,e)&=&\frac{n^4}{32}\left[J_{n-2}(ne)-2eJ_{n-1}(ne)
+\frac{2}{n}J_n(ne)+2eJ_{n+1}(ne)-J_{n+2}(ne)\right]^2 \nonumber \\
&+&\frac{n^4}{32}(1-e^2)\left[J_{n-2}(ne)-2J_n(ne)+J_{n+2}(ne)\right]^2 + \frac{n^2}{24}\left[J_n(ne)\right]^2.
\label{g(n,e)}
\ea
Then the orbit-averaged total gravitational wave power in all harmonics is
\bb
P = \frac{32}{5}\frac{G^4}{c^5}\frac{M^5\eta^2}{a^5}\sum_{n=1}^\infty g(n,e) = \frac{32}{5}\frac{G^4}{c^5}\frac{M^5\eta^2}{a^5}
\frac{1+\frac{73}{24}e^2+\frac{37}{96}e^4}{(1-e^2)^{7/2}}.
\label{P}
\ee

Since each graviton has energy $\hbar\omega = n\hbar\sqrt{GM/a^3}$, the orbit-averaged number rate of graviton emission is
\bb
\frac{dN}{dt} = \frac{32}{5}\frac{G^{7/2}}{\hbar c^5}\frac{M^{9/2}\eta^2}{a^{7/2}}\sum_{n=1}^\infty \frac{g(n,e)}{n}.
\label{dN/dt}
\ee
However, in this case there does not seem to be an explicit closed-form elementary function of the eccentricity $e$ for the infinite sum of $g(n,e)/n$ over the harmonic index $n$.

By comparison, the angular momentum $J$ per time radiated into gravitational waves (decreasing the orbital angular momentum $L$ of the masses by the same amount), averaged over one orbit, is given by Peters \cite{Peters:1963ux,Peters:1964qza} as
\bb
\frac{dJ}{dt} =
\frac{32}{5}\frac{G^{7/2}}{c^5}\frac{M^{9/2}\eta^2}
{a^{7/2}}
\frac{1+\frac{7}{8}e^2}{(1-e^2)^2}.
\label{torque}
\ee

Therefore, the ratio of the sum of magnitudes of the spin angular momenta $2\hbar$ to the magnitude of the total angular momentum vector emitted per orbit is
\bb
F(e) \equiv 2\hbar \frac{dN}{dJ} = \frac{2(1-e^2)^2}{1+\frac{7}{8}e^2}\sum_{n=1}^\infty \frac{g(n,e)}{n}.
\label{dN/dJ}
\ee
One can easily see that for circular orbits, with eccentricity $e=0$, in Eq.\ (\ref{g(n,e)}) only $J_0(0)$ is nonzero, and its value of 1 makes $g(n,0)$ nonzero only for $n=2$, giving $g(2,0)=1$ and hence $2\hbar dN/dJ = 1$ for $e=0$.  This is the case in which each graviton has angular momentum $2\hbar$ parallel to the orbital angular momentum $L$.

\section{High Eccentricity Graviton Number Emission}

Although I do not know how to get an explicit closed-form elementary function for $2\hbar dN/dJ$ as a function of the eccentricity $e$ for $0 < e < 1$, one can find it in the limit of unit eccentricity.  For simplicity, define
\bb
f \equiv 1-e^2.
\label{f}
\ee
Then for $f \ll 1$, the sum over $n$ of $g(n,e)/n$ is dominated by terms with very large $n$.

One can then use Eq. (8.455.1) on page 913 of Gradshteyn and Ryzhik \cite{GR}, that for real $x$ and integer $n \gg 1$,
\bb
J_n(x) \approx \frac{1}{\pi}\sqrt{\frac{2(n-x)}{3x}}\,
K_{\frac{1}{3}}\left\{\frac{[2(n-x)]^{\frac{3}{2}}}{3\sqrt{x}}\right\}.
\label{Jn(x)approx}
\ee
With $x=ne=n\sqrt{1-f}\approx n(1-f/2)$ for $f = 1-e^2 \ll 1$, this becomes
\bb
J_n(ne) \approx \frac{1}{\pi}\sqrt{\frac{f}{3}}\,
K_{\frac{1}{3}}\left(\frac{n}{3}f^{\frac{3}{2}}\right).
\label{Jn(ne)approx}
\ee
By using the recursion relation for the derivative of the modified Bessel functions $K_\nu(z)$, namely $K'_\nu(z) = -K_{\nu-1}(z)-(\nu/z)K_\nu(z)$, where the prime means the derivative with respect to its argument $z = (n/3)f^{3/2}$, and the fact that $K_\nu(z) = K_{-\nu}(z)$, one can get the derivative of the ordinary Bessel function with respect to its argument $x=ne$, $J'_n(x) = dJ_n(x)/dx$, as
\bb
J'_n(ne) \approx \frac{1}{\pi}\frac{f}{\sqrt{3}}
K_{\frac{2}{3}}\left(\frac{n}{3}f^{\frac{3}{2}}\right).
\label{Jn'(ne)approx}
\ee

Next, we can follow the use by Peters and Mathews \cite{Peters:1964zz} of the recurrence relations and Bessel's equation to rewrite $g(n,e)$ as their Eq.\ (A1), with the correction by Nikishov \cite{N} that the factor $(4/e^2)^2$ in the 3rd line should instead be $(4/e)^2$, to get
\ba
2e^4\frac{g(n,e)}{n} &=& (1-e^2)^3 n^3 J^2_n(ne) 
+ (1-e^2+\frac{1}{3}e^4) n J^2_n(ne) \nonumber \\
&-& e(4-3e^2)(1-e^2) n^2 J_n(ne) J'_n(ne) \nonumber \\
&+& e^2(1-e^2)^2 n^3 J'^2_n(ne)
+ e^2(1-e^2) n J'^2_n(ne).
\label{2e^4(g(n,e)/n}
\ea
Therefore, to evaluate the sum over $n$ of $g(n,e)/n$, we need to evaluate the sums of the five series that are each a power of $n$ multiplied by a bilinear in the corresponding Bessel function or its derivative.  When the powers of $n$ are one greater, Peters and Mathews \cite{Peters:1964zz} were able to evaluate the corresponding five sums explicitly as $(1-e^2)^{-13/2}$ multiplied by finite polynomials of $e$ in their Eqs.\ (A3).  However, their procedure does not work when the exponents of $n$ are those of Eq.\ (\ref{2e^4(g(n,e)/n}), each being one less than the corresponding exponent for each sum evaluated by Peters and Mathews.

Nevertheless, for the leading behavior for the sum of $g(n,e)/n$ for small $f=1-e^2$, we can use the approximations of Eqs.\ (\ref{Jn(ne)approx}) and (\ref{Jn'(ne)approx}).  Since for $f \ll 1$ the sums are dominated by terms with large values of $n$ and do not change rapidly with $n$, we can replace the sums by integrals over $n$ to get the leading terms for $f \ll 1$.  The integrals are given by Eq.\ (6.576.4) on page 676 of Gradshteyn and Ryzhik \cite{GR}, which for equal arguments for the two modified Bessel functions and for integer exponent $p$ gives
\ba
\int_0^\infty x^p K_\mu(ax)K_\nu(ax)dx &=& \frac{2^{p-2}}{p!a^{p+1}}
\Gamma\left(\frac{p+1+\mu+\nu}{2}\right)
\Gamma\left(\frac{p+1-\mu-\nu}{2}\right) \nonumber \\
&\times&\Gamma\left(\frac{p+1+\mu-\nu}{2}\right)
\Gamma\left(\frac{p+1-\mu+\nu}{2}\right).
\label{KKint}
\ea

Letting $x=n$ and $a = (1/3)f^{3/2} = (1/3)(1-e^2)^{3/2} \ll 1$, employing the triplication formula for the gamma function, and using the approximations of Eqs.\ (\ref{Jn(ne)approx}) and (\ref{Jn'(ne)approx}) allows us to calculate that for $1-e^2 \ll 1$ and for positive integer exponents $p = 2k+1 > 0$ or $p = 2k >0$, the leading terms in powers of $1-e^2$ are
\ba
\sum_{n=1}^\infty n^{2k+1} J_n(ne) J_n(ne) 
&\approx& \frac{2^{2k}k!(3k+1)!}{3^{k+1/2}\pi(2k+1)!}(1-e^2)^{-3k-2}, \nonumber \\
\sum_{n=1}^\infty n^{2k} J_n(ne) J_n(ne) 
&\approx& \frac{(6k-1)!!}{2^{2k+1}3^k(2k)!!}(1-e^2)^{-3k-1/2}, \nonumber \\
\sum_{n=1}^\infty n^{2k+1} J_n(ne) J'_n(ne) 
&\approx& \frac{(6k+1)!!}{2^{2k+2}3^k(2k)!!}(1-e^2)^{-3k-3/2}, \nonumber \\
\sum_{n=1}^\infty n^{2k} J_n(ne) J'_n(ne) 
&\approx& \frac{2^{2k-1}k!(3k-1)!}{3^{k-1/2}\pi(2k)!}(1-e^2)^{-3k}, \nonumber \\
\sum_{n=1}^\infty n^{2k+1} J'_n(ne) J'_n(ne) 
&\approx& \frac{2^{2k}(3k+2)k!(3k)!}{3^{k+1/2}\pi(2k+1)!}(1-e^2)^{-3k-1}, \nonumber \\
\sum_{n=1}^\infty n^{2k} J'_n(ne) J'_n(ne) 
&\approx& \frac{(6k+1)(6k-3)!!}{2^{2k+1}3^k(2k)!!}(1-e^2)^{-3k+1/2}.
\label{generalsums}
\ea
One may note that
\bb
(3p-1)\sum_{n=1}^\infty n^p J'_n(ne) J'_n(ne) \approx 
(3p+1)(1-e^2)\sum_{n=1}^\infty n^p J_n(ne) J_n(ne).
\label{J'/J}
\ee

Then the leading terms for the sums with $n$, $n^2$, $n^3$, and $n^4$ become
\ba
\sum_{n=1}^\infty n J_n(ne) J_n(ne) 
&\approx& \frac{1}{\sqrt{3}\pi}(1-e^2)^{-2}, \nonumber \\
\sum_{n=1}^\infty n J_n(ne) J'_n(ne)
&\approx& \frac{1}{4}(1-e^2)^{-3/2}, \nonumber \\
\sum_{n=1}^\infty n J'_n(ne) J'_n(ne)
&\approx& \frac{2}{\sqrt{3}\pi}(1-e^2)^{-1}, \nonumber \\
\sum_{n=1}^\infty n^2 J_n(ne) J_n(ne) 
&\approx& \frac{5}{16}(1-e^2)^{-7/2}, \nonumber \\
\sum_{n=1}^\infty n^2 J_n(ne) J'_n(ne)
&\approx& \frac{2}{\sqrt{3}\pi}(1-e^2)^{-3}, \nonumber \\
\sum_{n=1}^\infty n^2 J'_n(ne) J'_n(ne)
&\approx& \frac{7}{16}(1-e^2)^{-5/2}, \nonumber \\
\sum_{n=1}^\infty n^3 J_n(ne) J_n(ne) 
&\approx& \frac{16}{3\sqrt{3}\pi}(1-e^2)^{-5}, \nonumber \\
\sum_{n=1}^\infty n^3 J_n(ne) J'_n(ne)
&\approx& \frac{35}{32}(1-e^2)^{-9/2}, \nonumber \\
\sum_{n=1}^\infty n^3 J'_n(ne) J'_n(ne)
&\approx& \frac{20}{3\sqrt{3}\pi}(1-e^2)^{-4}, \nonumber \\
\sum_{n=1}^\infty n^4 J_n(ne) J_n(ne) 
&\approx& \frac{1155}{256}(1-e^2)^{-13/2}, \nonumber \\
\sum_{n=1}^\infty n^4 J_n(ne) J'_n(ne)
&\approx& \frac{80}{3\sqrt{3}\pi}(1-e^2)^{-6}, \nonumber \\
\sum_{n=1}^\infty n^4 J'_n(ne) J'_n(ne)
&\approx& \frac{1365}{256}(1-e^2)^{-11/2}.
\label{sums}
\ea

As a check, the 4th, 6th, 8th, 10th, and 12th formulas above agree with the leading terms of what Peters and Mathews \cite{Peters:1964zz} calculated in their Eqs.\ (A3).  Then the 1st, 3rd, 5th, 7th, and 9th formulas above can be inserted into Eq.\ (\ref{2e^4(g(n,e)/n}) to give, again for $f \equiv 1-e^2 \ll 1$,
\bb
\sum_{n=1}^\infty \frac{g(n,e)}{n} \approx \frac{31}{6\sqrt{3}\pi}(1-e^2)^{-2}
\label{sum}
\ee
This can then be inserted into Eq.\ (\ref{dN/dJ}) to give, in the limit of unit eccentricity ($e=1$),
\bb
F(1) = 2\hbar \frac{dN}{dJ}(e=1) = \frac{248}{45\sqrt{3}\pi} \approx 1.012\,811\,600\,479.
\label{dN/dJlimit}
\ee

It is rather remarkable that even for eccentricity $e=1$ the number of gravitons emitted is so close to the angular momentum emitted divided by the spin angular momentum $2\hbar$ of each graviton.

\section{Independent Derivation for a Parabolic Orbit}

As a check of the derivation above, I used the independently derived energy spectrum for a parabolic orbit (eccentricity $e=1$) that has been given by Turner \cite{Turner1977}, by Berry and Gair \cite{Berry:2010gt}, and by Gr\"obner, Jetzer, Haney, Tiwari, and Ishibashi \cite{Grobner:2020fnb}, all of whom got results that I could show are equivalent to the simplified form below:

In terms of the frequency $f_c$ of a circular orbit of the same radius as the periapsis distance $r_p$, which would be $r_p = a(1-e)$ for an ellipical orbit with semimajor axis $a = r_p/(1-e)$, except that here I am taking the limit $e \rightarrow 1$ with fixed $r_p$, sending the semimajor axis to infinity,
\bb
f_c = \frac{1}{2\pi}\sqrt{\frac{GM}{r_p^3}},
\ee
the frequency $f$ of the radiation (both of these frequencies in cycles per second, and in this section $f$ is {\it not} $1-e^2$) and a conveniently normalized dimensionless frequency I shall call $x$ (which is the argument of each modified Bessel function of the second kind below, $K_{1/3}$ and $K_{2/3}$),
\bb
x = \sqrt{\frac{8}{9}}\frac{f}{f_c},
\ee
the gravitational wave energy $E$ per orbit and frequency interval is
\bb
\frac{dE}{df} = \frac{16}{5}\frac{G^3}{c^5}
\frac{M^4\eta^2}{r_p^2}
\left[(27x^4+x^2)K_{1/3}(x)^2-9x^3K_{1/3}(x)K_{2/3}(x)
+27x^4K_{2/3}(x)^2\right].
\ee

This then gives the number of gravitons emitted during one parabolic orbit of pariastron distance $r_p$ and orbital angular momentum $L = \sqrt{2GM^3\eta^2 r_p}$ as
\bb
\Delta N = \int\frac{dE}{hf} = \frac{1}{h}\int_0^\infty\frac{dE}{df}\frac{dx}{x}
= \frac{248}{15\sqrt{3}}\frac{G^3}{\hbar c^5}\frac{M^4\eta^2}{r_p^2}
= \frac{992}{15\sqrt{3}}\frac{G^5M^{10}\eta^6}{\hbar c^5 L^4}.
\label{DeltaN}
\ee

One can compare this with the calculations of Peters \cite{Peters:1964qza,Peters:1964zz} for the angular momentum radiated during one parabolic orbit:
\bb
\Delta J = 6\pi\frac{G^3M^4\eta^2}{c^5 r_p^2} 
= 24\pi\frac{G^5M^{10}\eta^6}{c^5 L^4}.
\label{DeltaJ}
\ee
This then gives
\bb
F(1) = 2\hbar \frac{\Delta N}{\Delta J}(e=1) = \frac{248}{45\sqrt{3}\pi} \approx 1.012\,811\,600\,479.
\label{DeltaN/DelataJ}
\ee
for eccentricity $e=1$, just as we found before from the sum of $g(n,e)/n$ in the limit that $1-e^2$ was taken to zero.

\section{Graviton Numbers Emitted During Inspiral}

Next, we would like to get estimates of the total number of gravitons emitted during the inspiral from arbitrary initial orbital angular momentum $L_i$ and initial eccentricity $e_i$.  Since the fractional change in the orbital angular momentum and in the eccentricity during each orbit is small (unlike the case for the semimajor axis $a$, nonrelativistic orbit energy $E = - GM^2\eta/(2a)$, and orbital period $\tau = 2\pi\sqrt{a^3/(GM)}$, which can have large fractional changes during orbits of large eccentricity \cite{Page:2024rsv,Page:2024zpr}), we can use Eq.\ (5.11) of \cite{Peters:1964zz} to get the ratio of the orbital angular momentum $L(e)$, when the eccentricity is $e$, to $L_1$, what it would have been at eccentricity $e=1$ if the inspiral had started there:
\bb
\lambda(e) \equiv \frac{L(e)}{L_1} = e^{6/19}\left(\frac{304+121e^2}{425}\right)^{435/2299}
= (1-f)^{3/19}\left(1-\frac{121}{425}f\right)^{435/2299},
\label{lambda}
\ee
where now we return to $f \equiv 1-e^2$.  Then if an inspiral starts at initial orbital angular momentum $L_i$ and eccentricity $e_i$, the orbital angular momentum during the inspiral as the eccentricity decreases below $e_i$ is
\bb
L(e) = L_i\frac{\lambda(e)}{\lambda(e_i)}.
\label{L(e)}
\ee

In order to calculate the number of gravitons 
\bb
N(L_i,e_i) = [L_i/(2\hbar)]\nu(e_i)
\label{N(Le)}
\ee
emitted during the inspiral from initial orbital angular momentum $L_i$ and eccentricity $e_i$ to final eccentricity very close to zero (assuming the initial orbit is highly nonrelativistic) and final angular momentum much smaller than the initial value, ideally one would need
\bb
F(e) = 2\hbar \frac{dN}{dJ}(e).
\label{F}
\ee
Then one could calculate the number of gravitons emitted in going from initial orbital angular momentum $L_i$ at initial eccentricity $e_i$ to final eccentricity indistinguishable from 0 as
\bb
N(L_i,e_i) \equiv \frac{L_i}{2\hbar}\nu(e_i) = \frac{L_i}{2\hbar\lambda(e_i)}\int_0^{e_i} F(e) \frac{d\lambda}{de} de,
\label{N2}
\ee
or
\bb
\nu(e_i) \equiv \frac{2\hbar N(L_i,e_i)}{L_i} = \int_0^{e_i} \frac{F(e)}{\lambda(e_i)} \frac{d\lambda}{de} de
         = \int_0^{\lambda(e_i)} \frac{F(e(\lambda))}{\lambda(e_i)} d\lambda,
\label{nu}
\ee
which is the sum of the spin angular momentum magnitudes of the gravitons emitted ($2\hbar$ times the number $N$ of gravitons emitted) divided by the magnitude $J = L_i$ of the total angular momentum vector emitted in the gravitational radiation during nonrelativistic inspiral from an initial eccentricity $e_i$ down to a final eccentricity indistinguishable from zero when the orbit becomes relativistic and then rapidly leads to merger, with its final angular momentum being assumed to be much less than the initial orbital angular momentum $L_i$.

I have not been able to find an easily evaluated precise formula for $F(e)$ or for $F(e(\lambda))$, and I could not get Mathematica to get good convergence for the sum over $n$ for $g(n,e)/n$ for $e > 0.6$, because of the large number of terms contributing significantly, but for $e \leq 0.6$, I found $F(e)$ could be fit well by this simple linear function of the squared eccentricity:
\bb
F(e) \approx 1 
+ \left(\frac{248}{45\sqrt{3}\pi}-1\right)e^2 
\approx 1 + 0.012\,811\,600\, e^2.
\label{Fapprox}
\ee
To get a corresponding approximation for $F(e(\lambda))$, I found that an approximate inversion for Eq.\ (\ref{lambda}) for $\lambda(e)$ that fits well near both $\lambda=0$ and $\lambda=1$ with
\bb
g \equiv \lambda^{19/3} = e^2\left(\frac{304+121e^2}{425}\right)^{145/121}
= (1-f)\left(1-\frac{121}{425}f\right)^{145/121}.
\label{g}
\ee
(this $g$ not to be confused with $g(n,e)$, and using, as nearly always, $f \equiv 1-e^2$) is
\bb
e^2\approx 1.494g - 1.065g^2 + 1.102g^3 - 0.731g^4 +0.200g^5,
\ee
where 1.494 is an approximation for $(425/304)^{145/121} \approx 1.494\,093\,9064$, 1.065 is an approximation for 
$(29/85)(425/304)^{411/121} \approx 1.064\,756\,2736$, and the other coefficients are approximations for explicit rational fractions plus positive or negative integer multiples of the first two coefficients that are not overly illuminating, so I shall not bother listing their precise values.

Inserting this $e^2(\lambda)$ into $F(e) \approx 1 + 0.012811 e^2$ from Eq.\ (\ref{Fapprox}) and then using this approximate $F(e(\lambda))$ in Eq.\ (\ref{nu}) leads to (with $g_i = g(e_1)$ in terms of the initial eccentricity $e_1$)
\ba
\nu(e_i) &\!\!\!\approx\!\!\!&1\!+\!(0.01281)
\left(\frac{1.494}{22/3}g_i\!-\!\frac{1.065}{41/3}g_i^2\!+\!\frac{1.102}{60/3}g_i^3\!-\!\frac{0.731}{79/3}g_i^4\!+\!\frac{0.200}{98/3}g_i^5\right)
\nonumber \\
&\!\!\!\approx\!\!\!& 1 + 0.002\,610 g_i -0.000\,998 g_i^2 + 0.000\,706 g_i^3 - 0.000\,356 g_i^4 + 0.000\,079 g_i^5
\nonumber \\
&\!\!\!\approx\!\!\!& 1.002\,041 - 0.001\,701 (1-g_i) - 0.000\,230 (1-g_i)^2 - 0.000\,068 (1-g_i)^3 
\nonumber \\
&\!\!\!+\!\!\!& 0.000\,037 (1-g_i)^4 - 0.000\,079 (1-g_i)^5.
\label{nulinear}
\ea
I should emphasize that since I do not know even the approximate behavior of $F(e)$ for $0.6 < e < 1$, though I would assume that it is a highly smooth function as it appears to be for $0 < e < 0.6$, I do not know how many of the digits in the formulas above are correct, so they are mainly suggestive.  

Another set of estimates that probably are more accurate for small eccentricity $e$ uses the fact that for such $e \ll 1$, one can use the power series for the Bessel functions in Eq.\ (\ref{g(n,e)}), for which only those for $n \leq 5$ give terms with powers of $e$ up through $e^4$.  In particular, one gets
\ba
g(1,e) &=& \frac{29}{192} e^2 - \frac{119}{768} e^4 + O(e^6), \nonumber \\
g(2,e) &=& 1 - 5 e^2 + \frac{55}{6} e^4 + O(e^6), \nonumber \\
g(3,e) &=& \frac{729}{64} e^2 - \frac{13851}{256} e^4 + O(e^6), \nonumber \\
g(4,e) &=& 64 e^4 + O(e^6).
\label{4g(n,e)}
\ea

One can check that these terms give
\bb
\sum_{n=1}^\infty g(n,e) = 1 + \frac{157}{24} e^2 + \frac{605}{32} e^4 + O(e^6),
\label{sumg(n,e)}
\ee
which agrees through this order $e^4$ with the exact closed-form formula of the last equation (unnumbered) in the Appendix of Peters and Mathews \cite{Peters:1964zz}:
\bb
\sum_{n=1}^\infty g(n,e) = (1-e^2)^{-7/2}\left(1 + \frac{73}{24} e^2 + \frac{37}{96} e^4\right).
\label{PMsumg(n,e)}
\ee

For the case in which no closed-form formula is known, these terms give
\bb
\sum_{n=1}^\infty \frac{g(n,e)}{n} = \frac{1}{2} + \frac{139}{96} e^2 + \frac{919}{384} e^4 + O(e^6),
\label{sumg(n,e)/n}
\ee
Then one can use Eq.\ (\ref{dN/dJ}) to get
\bb
F(e) \equiv 2\hbar \frac{dN}{dJ} = \frac{2(1-e^2)^2}{1+\frac{7}{8}e^2}\sum_{n=1}^\infty \frac{g(n,e)}{n}
     = 1 + \frac{1}{48} e^2 - \frac{3}{128} e^4 + O(e^6).
\label{seriesdN/dJ}
\ee

Since Eq.\ (\ref{dN/dJlimit}) gives at unit eccentricity ($e=1$) the value $F(1) = 248/(45\sqrt{3}\pi) \approx 1.012\,811\,600$,
a one-parameter set of approximations for $F(e)$ for all $0 \leq e \leq 1$ that for an appropriate value of the parameter $s$ seems as if it should be better than the approximation linear in $e^2$ given in Eq.\ (\ref{Fapprox}) is
\bb
F(s,e) \equiv 1 + \frac{1}{48} e^2 - \frac{3}{128} e^4 + C e^{2s}
\approx F(e) \equiv 2\hbar \frac{dN}{dJ} = \frac{2(1-e^2)^2}{1+\frac{7}{8}e^2}\sum_{n=1}^\infty \frac{g(n,e)}{n}
\label{approxdN/dJ}
\ee
with
\bb
C = \frac{248}{45\sqrt{3}\pi} - \frac{383}{384} \approx 0.015\,415\,767\,147,
\label{C}
\ee
with this $F(s,e)$ for $s \geq 3$ agreeing with Eq.\ (\ref{seriesdN/dJ}) that is valid for $e \ll 1$ and also, with the inclusion of the last term ($Ce^{2s}$), fits the exact value at $e=1$.  The uncertainty for eccentricities that are neither very small nor equal to 1 is partially modeled by the uncertainty in the exponent $2s$, which I would expect fits best for some $2s \geq 6$ (such as $2s=6$, which would approximate $F(e)$ by a cubic in $e^2$) so that Eq.\ (\ref{approxdN/dJ}) agrees with Eq.\ (\ref{seriesdN/dJ}) for $e \ll 1$.

Now we can apply integration by parts to Eq.\ (\ref{nu}) with the unknown exact $F(e)$ replaced by each explicit approximate $F(s,e)$ to get
\ba
&&\nu(e) \approx \nu(s,e_i) = F(s,e_i) - \int_0^{e_i} \frac{\lambda(e)}{\lambda(e_i)} \frac{dF(s,e)}{de} de 
\nonumber \\
&&=\!F(s,e_i)\!-\!e_i^{-\frac{6}{19}}\!\left(1\!+\!\frac{121}{304}\!e_i^2\!\right)\!^{-\frac{435}{2299}}\!\int_0^{e_i} \!de e^\frac{6}{19}\!\left(1\!+\!\frac{121}{304}\!e^2\right)^\frac{435}{2299}\!\left(\frac{e}{24}\!-\!\frac{3e^3}{32}\!+\!2sCe^{2s-1}\!\right)\!. \nonumber \\
&&
\label{nu2}
\ea
To give an explicit approximation for this integral that depends on the parameter $s$ in the last term of $F(s,e)$, I shall use the following fit for the nonpolynomial factor
that matches its value and derivative at $e^2 = 0$ and its value at $e^2 = 1$:
\bb
\left(1+\frac{121}{304}e^2\right)^{\frac{435}{2299}} 
\approx 1 + \frac{435}{5776} e^2 - \left[\frac{6211}{5776}-\left(\frac{425}{304}\right)^{\frac{435}{2299}}\right] e^4
\equiv 1 + A e^2 - B e^4.
\label{polyapprox}
\ee
Then one gets
\ba
(1+Ae_i^2-Be_i^4)\nu(s,e_i) &\approx& 1+\left(\frac{1}{352}+A\right)e_i^2 + \left(-\frac{9}{5248}+\frac{11}{984}A-B\right)e_i^4 \nonumber \\
&&+ \left(-\frac{11}{1280}A-\frac{41}{2880}B\right) e_i^6 + \left(\frac{123}{10112}B\right)e_i^8 \nonumber \\
&&+ \frac{3Ce_i^{2s}}{3+19s} + \frac{22ACe_i^{2s+2}}{22+19s} - \frac{41BCe_i^{2s+4}}{41+19s}.
\label{nu3}
\ea

Since the quantity inside the square brackets in Eq.\ (\ref{polyapprox}), $B \approx 0.009\,860\,952\,250$, is less than 1\% and is not a rational number, it is convenient to drop it and use $A = 435/5776 \approx 0.075\,311\,634\,349$ to get rational coefficients for the integer powers of $e_i^2$ that do not have $s$ in the exponent and the factor $C$ given in Eq.\ (\ref{C}):
\ba
\nu(s,e_i) \approx \left(1+\frac{435}{5\,776}e_i^2\right)^{-1}&&\!\!\!\!\!\!\!\!\!\!
\left[1 + \frac{9\,931}{127\,072}e_i^2 - \frac{827}{947\,264} e_i^4 - \frac{957}{1\,478\,656} e_i^6\, + \right. \nonumber \\
+ && \!\!\!\!\!\!\!\!\!\!\! \left.\left(\frac{248}{45\sqrt{3}\pi} - \frac{383}{384}\right)\!\! 
\left(\frac{3e_i^{2s}}{3+19s} + \frac{435}{5\,776}\frac{22 e_i^{2s+2}}{22+19s} \right)\!\right]\!\!.
\label{nu4}
\ea

A quantity of particular interest is $\nu(1)$, the spin angular momentum of each graviton ($2\hbar$) multiplied by the number $N$ of gravitons emitted and divided by the magnitude $J$ of the total angular momentum emitted in gravitons when the orbit starts off with unit eccentricity ($e_1 = 1$, an essentially parabolic initial orbit), and inspirals to negligible eccentricity, $e_f \ll 1$.  The approximate Eq.\ (\ref{nu4}) above evaluated at $e_1 = 1$ gives
\ba
\nu(s,1) &\approx& \frac{717\,977\,929}{717\,097\,216} + \left(\frac{248}{45\sqrt{3}\pi} 
- \frac{383}{384}\right) \frac{409\,926 + 511\,062 s}{6\,211(3+19s)(22+19s)}, \nonumber \\
\nu(0,1) &\approx& \frac{717\,977\,929}{717\,097\,216} + \left(\frac{248}{45\sqrt{3}\pi} 
- \frac{383}{384}\right) 1  \approx 1.016\,643\,931\,168, \nonumber \\
\nu(1,1) &\approx& \frac{717\,977\,929}{717\,097\,216} + \left(\frac{248}{45\sqrt{3}\pi} 
- \frac{383}{384}\right)\frac{460\,494}{2\,801\,161} \approx 1.003\,762\,423\,307, \nonumber \\
\nu(2,1) &\approx& \frac{717\,977\,929}{717\,097\,216} + \left(\frac{248}{45\sqrt{3}\pi} 
- \frac{383}{384}\right)\frac{47\,735}{509\,302} \approx 1.002\,673\,027\,078, \nonumber \\
\nu(3,1) &\approx& \frac{717\,977\,929}{717\,097\,216} + \left(\frac{248}{45\sqrt{3}\pi} 
- \frac{383}{384}\right)\frac{161\,926}{2\,453\,345} \approx 1.002\,245\,637\,515, \nonumber \\
\nu(4,1) &\approx& \frac{717\,977\,929}{717\,097\,216} + \left(\frac{248}{45\sqrt{3}\pi} 
- \frac{383}{384}\right)\frac{1\,227\,087}{24\,042\,781} \approx 1.002\,014\,948\,522, \nonumber \\
\nu(5,1) &\approx& \frac{717\,977\,929}{717\,097\,216} + \left(\frac{248}{45\sqrt{3}\pi} 
- \frac{383}{384}\right)\frac{494\,206}{11\,869\,221} \approx 1.001\,870\,039\,727, \nonumber \\
\nu(6,1) &\approx& \frac{717\,977\,929}{717\,097\,216} + \left(\frac{248}{45\sqrt{3}\pi} 
- \frac{383}{384}\right)\frac{579\,383}{16\,471\,572} \approx 1.001\,770\,409\,376, \nonumber \\
\nu(\infty,1) &\approx& \frac{717\,977\,929}{717\,097\,216} \approx 1.001\,228\,164\,021.
\label{nu(s,1)}
\ea

Using $s \leq 2$ would not give $F(s,e)$ agreeing with its low-eccentricity form $F(e) = 1 + e^2/48 -3e^4/128 + O(e^6)$, though that fact does not prove that $s > 2$ is needed to give the best estimate for $\nu(e)$ for large $e$, such as $\nu(1)$.  In the absence of further information, I would tend to prefer $s=3$ so that $F(3,e) = 1+e^2/48-3e^4/128+Ce^6$ given by Eq.\ (\ref{approxdN/dJ}) with $C$ given by Eq.\ (\ref{C}) is a cubic polynomial in $e^2$ that agrees with $F(e) = 1 + e^2/48 -3e^4/128 + O(e^6)$.  Inserting this cubic polynomial into Eq.\ (\ref{nu2}), using the exact $\lambda(e)$ from Eq.\ (\ref{lambda}) (without the polynomial approximation of Eq.\ (\ref{polyapprox}) and the truncation of dropping the $Be^4$ term that were used to get the approximate values of $\nu(s,1)$ in Eq.\ (\ref{nu(s,1)})), and evaluating the integral by Mathematica led to
\bb
\nu(3,1) = 1.002\,189\,127\,062.
\label{nu(3,1)}
\ee
Note that the approximate Eq.\ (\ref{nu2}) with $B$ dropped gave $\nu(3,1) \approx 1.002\,245\,637\,515$ above, which is about 1.000\,056\,387 times the `exact' $\nu(3,1)$ (using $F(3,e) = 1+e^2/48-3e^4/128+Ce^6$, which of course is only a plausibly best simple guess for an approximation for the unknown $F(e)$), and the `exact' $\nu(3,1)$ is also not too far from the simpler approximation of Eq.\ (\ref{nulinear}), using the approximation $F(e) \approx 1 + 0.012811 e^2$ that is linear in $e^2$, which gives $\nu(1) \approx 1.002\,041$.

One attempt to get upper and lower bounds on $\nu(e=1)$ would be to assume reasonable upper and lower bounds on $F(e)$.  Since for $e \leq 1$ it has the form $F(e) = 1 + e^2/48 -3e^4/128 + O(e^6)$, and since $1+1/48 > F(1) = 248/(45\sqrt{3}\pi) \approx 1.0128 > 1 + 1/48 - 2/128$, it seems highly plausible that $1 + e^2/48 > F(e) > 1 + e^2/48 -3e^4/128$ for all $0 \leq e \leq 1$, and even more plausible that $\nu(1) = \int_0^1 F d\lambda$ is between the values given by replacing $F$ with $1 + e^2/48$ and with $1 + e^2/48 -3e^4/128$ respectively.  These two integrals were evaluated by Mathematica and gave $1.003\,327\,748\,211$ and $1.001\,219\,302\,812$ respectively, so I would highly expect the true value to lie between those two conservative limits.  I also suspect that if the actual $\nu(1)$ were rounded to 4 digits, it would be 1.002, but this would need to be confirmed by further calculations.  However, it seems quite clear that the number of gravitons emitted during binary inspiral is well within 1\% of the magnitude of the total angular momentum emitted divided by the spin angular momentum of a single graviton.

\section{Conclusions}

For two black holes (or other effectively point masses) inspiraling from an initially highly nonrelativistic Keplerian orbit (e.g., initial periapsis distance of closest approach much greater than the Schwarzschild radius of the total mass) with initial orbital angular momentum $L_i$ and eccentricity $e_i$, approximate formulas are given for the spin angular momentum of the graviton ($2\hbar$) multiplied by the number $N$ of gravitons emitted and divided by the magnitude $J$ of the angular momentum of the gravitational radiation, both at the differential level ($F(e) = 2\hbar dN/dJ$) for an infinitesimal change in the eccentricity, and at the total level ($\nu(e_i) = 2\hbar N/L_i$) for the total graviton number $N$ emitted during inspiral to negligible final eccentricity from initial orbital angular momentum $L_i$ and eccentricity $e_i$.  It is calculated that at unit eccentricity, $F(1) = 248/(45\sqrt{3}\pi) \approx 1.012\,811\,600\,479$ (so that the number of gravitons emitted near this eccentricity is a bit less than 1.3\% more than the angular momentum emitted divided by the spin angular momentum of the graviton, $2\hbar$).  A precise formula for $\nu(e_i)$ was given only for initial eccentricities very small ($e_i \ll 1$), but it appears that the total number of gravitons emitted during the total inspiral is always less than about 1.002 times the angular momentum emitted divided by $2\hbar$.

\section{Note Added}

After this paper was posted on the arXiv, Youngjoo Chung and Hyun Seok Yang \cite{Chung:2024ymj} extended the results in a tour de force to find alternative exact infinite series for $F(e)$ and $\nu(e_i)$ which did not have Bessel functions in them but instead have (rather complicated) finite sums of rational numbers multipliying the even powers of the eccentricity $e$ in a convergent series expansion for $F(e)$, and an infinite sum of fractional powers of $e$ multiplying hypergeometric functions of $-121e_i^2/304$ for $\nu(e_i)$.  Unlike my failure to be able numerically to sum the Bessel function series for $F(e)$ for $e > 0.6$, they were able to sum their series for both $F(e)$ and $\nu(e_i)$ to high precision for any $e$ between 0 and 1 inclusive and find $\nu(1) = 1.002\,268\,666\,2$, which is 1.000\,079\,365 times my `best guess' $\nu(3,1) = 1.002\,189\,127\,062$ in Eq.\ (\ref{nu(3,1)}) above, or about one part in 12\,600 larger than my estimate.

\section*{Acknowledgments}

This work was motivated by research by Ashaduzzaman Joy on the number of gravitons emitted by black-hole coalescences observed by LIGO and Virgo (paper in preparation).  Hyun Seok Yang also kindly pointed out typos in the original form of my Eqs. (\ref{Jn(ne)approx}), (\ref{Jn'(ne)approx}), (\ref{DeltaJ}), and (\ref{4g(n,e)}), and in three other places in my text, which I have now replaced with his corrections.  Financial support was provided by the Natural Sciences and Engineering Research Council of Canada.

\end{document}